\documentclass[aps,prc,twocolumn,groupedaddress]{revtex4-2}

\usepackage{graphicx}
\usepackage{slashed}
\usepackage{color}
\usepackage{amsmath}
\usepackage{amsfonts}
\usepackage[normalem]{ulem}
\newcommand{\nopieft}{\mbox{$\slashed{\pi}$}EFT }
\newcommand{\nopi}{\mbox{$\slashed{\pi}$}}

\newcommand{\be}{\begin{equation}}
\newcommand{\ee}{\end{equation}}

\newcommand{\bra}{\langle}
\newcommand{\ket}{\rangle}

\newcommand{\bea}{\begin{eqnarray}}
\newcommand{\eea}{\end{eqnarray}}
\newcommand{\nn}{\nonumber}

\newcommand{\invfm}{\ensuremath{\,\mathrm{fm}^{-1}}}
\newcommand{\reff}{\ensuremath{r_{\mathrm{eff}}}}
\newcommand{\fm}{\,\mathrm{fm}}
\newcommand{\AS}{\mathcal{A}}
\newcommand{\rv}{\ensuremath{\mathbf{r}}}
\newcommand{\nvec}{\ensuremath{\boldsymbol{n}}}
\newcommand{\mvec}{\ensuremath{\boldsymbol{m}}}
\newcommand{\rvec}{\ensuremath{\boldsymbol{r}}}
\newcommand{\xvec}{\ensuremath{\boldsymbol{x}}}
\newcommand{\yvec}{\ensuremath{\boldsymbol{y}}}
\newcommand{\zvec}{\ensuremath{\boldsymbol{z}}}
\newcommand{\dvec}{\ensuremath{\boldsymbol{d}}}

\begin{document}
%============================================================
\title{Spectrum of light nuclei in a finite volume}

\author{R. Yaron}
\affiliation{The Racah Institute of Physics,The Hebrew University, Jerusalem 9190401, Israel}

\author{B. Bazak}
\email{betzalel.bazak@mail.huji.ac.il}
\affiliation{The Racah Institute of Physics,The Hebrew University, Jerusalem 9190401, Israel}

\author{M. Sch\"{a}fer}
\email{schafer.martin@mail.huji.ac.il}
\affiliation{The Racah Institute of Physics,The Hebrew University, Jerusalem 9190401, Israel}

\author{N. Barnea}
\email{nir@phys.huji.ac.il}
\affiliation{The Racah Institute of Physics,The Hebrew University, Jerusalem 9190401, Israel}

\date{\today}

%===============
\begin{abstract}
%===============
Lattice quantum chromodynamics calculations of multi-baryon systems with physical quark masses would start a new age of \emph{ab initio} predictions in nuclear physics.
Performed on a finite grid, such calculations demand extrapolation of their finite volume numerical results to free-space physical quantities.
Such extraction of the physical information can be carried out fitting effective field theories (EFTs) directly to the finite-volume results or utilizing the L\"uscher free-space formula or its generalizations for extrapolating the lattice data to infinite volume. 
To understand better the effect of periodic boundary conditions on the binding energy of few nucleon systems we explore here light nuclei with physical masses in a finite box and in  free space. 
The stochastic variational method is used to solve the few-body systems. Substantial optimizations of the method are introduced to enable efficient calculations in 
a periodic box. With the optimized code, we perform accurate calculations of light nuclei $ A\le 4$ within leading order pionless EFT.
Using L\"uscher formula for the two-body system, and its generalization for 3- and 4-body systems, we examine the box effect and explore possible limitations of these formulas for the considered nuclear systems.
\end{abstract}
\maketitle

%=====================
\section{Introduction}
%=====================
Nuclear interaction and the corresponding properties of nuclei are low energy manifestations of quantum chromodynamics (QCD), the underlying theory of the strong force. While QCD is perturbative at high energies, at low energies, relevant for nuclear physics, it is non-perturbative. To obtain \emph{ab initio} QCD predictions of few-nucleon systems one must resort to rather demanding and complex numerical calculations on a finite lattice, known as lattice QCD (LQCD). Such calculations are performed with a discrete spacing between lattice sites and in a finite volume with periodic, or twisted, boundary conditions. For infinite box size and infinitesimal spacing, LQCD provides an exact solution of QCD \cite{QCD}.

% which lattice QCD calculations have been done so far and what is the lowest mass achieved
One of the main challenges here is to calculate LQCD prediction for nuclear physics using physical quark masses which translates to the physical pion mass $m_\pi \approx 140$~MeV. Lattice calculations for higher $m_\pi$ were performed by several groups: the HAL QCD collaboration \cite{HALQCD11,HALQCD12,HALQCD18,HALQCD19,HALQCD20,HALQCD21} extracts the free-space nuclear potential directly from finite volume results. 
Other collaborations \cite{PACS12,PACS15,NPLQCD13,NPLQCD15,NPLQCD18,NPLQCD21,NPLQCD21b,NPLQCD21c,NPLQCD22,CALLAT17,clalat21,MAINZ19,MAINZ21} provide finite volume quantities which must be extrapolated to free space in order to obtain corresponding physical values. We note that the first LQCD results with reasonable signal-to-noise ratio near the physical $m_\pi$  have been introduced quite recently \cite{HALQCD19,HALQCD20,HALQCD21}.

% Lucher formalism, Sebastian, Maissner, ...
For the two-body system, the implications of finite volume effects on the spectrum are well understood through the L\"uscher formalism \cite{Lucher, Lucher2}, under the limitations $R \ll \kappa^{-1} \ll L$ where $R$ is the potential range, $\kappa$ is the binding momentum, and $L$ is the box size. For a physical $m_\pi$ the long-range part of the nuclear interaction has $R \approx 2\fm$, while for the deuteron $1/\kappa \approx 5\fm$. In LQCD calculations the box size $L$ is limited due to computational costs and currently is typically taken to be around $5\fm$; the fit to the L\"uscher formalism is therefore questionable. As the box becomes smaller, one needs to take also corrections to L\"uscher's asymptotic formula, see, for example, Refs.~\cite{KL:corrections2,KL:corrections3,KL:corrections}. 
Theoretical advances beyond L\"uscher's two-body formalism are focused predominantly on a three-body system \cite{BriDav13,HanSha14,MeiRioRus15,BriHanSha17,HamPanRus17}. 
Recently, it has been pointed out, that finite-volume effects in few-body bound states are to a large extent governed by the asymptotic part of their wave function which describes separation into two clusters with the lowest threshold \cite{KonLee18,Kon20}.    

%EFT theories and methods solving few-body Schrodinger equation with a periodic boundary condition
A different way to approach finite-volume effects was introduced in Ref.~\cite{EliBazBar20}. There it was suggested that effective field theories (EFTs) can be used to extrapolate
LQCD results to infinite volume. In this approach the nuclear few-body problem is solved on the same finite volumes as the LQCD calculations and the EFT's low energy constants (LECs) are fitted to reproduce the LQCD results. The infinite-volume quantities are then obtained solving the EFT in free-space.  
Specifically, in Ref.~\cite{EliBazBar20} a baryonic EFT, aka pionless EFT (\nopi EFT) \cite{EFT1,EFT2,EFT3,EFT4}, with the nucleons as the only degrees of freedom, 
was applied to reproduce the nuclear spectrum obtained by the NPLQCD collaboration
at $m_\pi=806$ MeV \cite{NPLQCD13}. 
For applications see Refs.~\cite{EliBazBar20,DetSha21,NPLQCD21c}.  

%STOCHASTIC VARIATIONAL METHOD
Fixing the LECs using finite volume data demands an accurate solution of the few-body Schr\"odinger equation with a periodic boundary condition. The main obstacle is the interplay between three different scales -- the interaction range $R$, the size of the wave function, and the box size $L$. 
The correlated Gaussians based Stochastic Variational Method (SVM), which we employ here, 
was proven to be versatile enough to describe finite-volume effects \cite{YinBlu13} 
and a reliable tool to fix \nopieft  to LQCD data \cite{EliBazBar20}. 
Recently, a variant of this approach was proposed in \cite{MIT}, where the basis functions are optimized using differentiable programming rather than in a stochastic manner.

%OTHER METHODS
The few-body bound state problem with periodic boundary conditions was also addressed by other methods such as discrete variable representation \cite{DVR} or quantum Monte Carlo \cite{QMC}. Both approaches suffer from slow convergence in the case of a small interaction range $R$. The Monte Carlo techniques are further burdened by the sign problem in Fermionic systems. We note that the problem of slower convergence is also pertinent to SVM, however, it starts to be significant at smaller $R$ than in the two aforementioned methods. 

% what we do in this work
Current progress in nuclear LQCD suggests that first calculations for physical $m_\pi$ might be reachable in a foreseeable future. This raises an important question of benefits or possible limitations of extracting free-space quantities using the L\"{u}scher formalism or its few-body extension \cite{KonLee18}, with respect to EFT approach.   
Using the Stochastic Variational Method (SVM) for free-space and largely optimized SVM code with periodic box boundary conditions (Box-SVM) we study the physical $\rm ^2H$, $\rm ^3H$, and $\rm ^4He$ nuclear systems. 
The interaction between nucleons is described using leading order (LO) \nopieft fitted to experimental data. We perform accurate few-body calculations of energy shifts induced by different finite volumes. Our results are compared to the predictions of the L\"{u}scher formula and its generalization for $N > 2$. 

%overview of the paper
The paper is organized as follows. In the next section we shortly describe LO \nopi EFT. 
The L\"{u}scher formula and its few-body generalizations, as well as the asymptotic normalization coefficients (ANCs), are introduced in Section III. Numerical tools used to solve the few-body systems, SVM and Box-SVM, are described in Section IV. Our results are presented in Section V, followed by conclusions and outlook. Some technical points are described in the appendices. 

%====================
\section{LO \nopieft}
%====================
At leading order, the \nopieft potential contains both two-body as well as three-body contact interactions. Regulated with a momentum cutoff $\Lambda$, this interaction takes the form 
\begin{align}
\hat V &= \sum_{i<j} 
V_{S=1,T=0}(\rv_{ij}) + V_{S=0,T=1}(\rv_{ij}) \nn\\
&+\sum_{i<j<k} \sum_{cyc} V_3(\rv_{ij},\rv_{jk}), 
\end{align}
where $\rv_{ij}=\rv_{i}-\rv_{j}$ is the relative distance between particles $i$ and $j$,
\be 
    V_{S,T}(\rv_{ij})=C_{S,T}(\Lambda)\hat P_{S,T} G_\Lambda(\rv_{ij})\,,
\ee 
and
\be 
V_3({\bf r}_{ij},{\bf r}_{jk})=D_1(\Lambda)\hat P_{S=1/2,T=1/2}G_\Lambda({\bf r}_{ij})G_\Lambda({\bf r}_{jk})\,.
\ee
Here $\hat P_{S,T}$ is a projection operator on the spin-isospin channel $S,T$, 
$G_\Lambda(\rv)$ is a smeared delta function 
$$G_\Lambda(\rv)=\frac{\Lambda^3}{8\pi^{3/2}}\exp[-\Lambda^2 r^2/4]\,,$$  
and $C_{S=1,T=0},C_{S=0,T=1},D_1$ are the theory's LECs.

The two-body spin-triplet and spin-singlet LECs $C_{S=1,T=0},C_{S=0,T=1}$ were fitted to the deuteron binding energy $B_d=2.2246$ MeV \cite{deuteronBE} and to the neutron-neutron spin-singlet scattering length $a_0=-18.95\fm$ \cite{nnSCTlngth1,nnSCTlngth2}, respectively. 
The three-body LEC was fitted to the triton binding energy, $B_t=8.482$ MeV \cite{tritonBE}. Calculations were done for cutoff values $\Lambda\in[1,10]\invfm.$
More details about the EFT and the fitting procedure are given in Refs.~\cite{BazEliKol16,SchBazBar21}.

%===============================
\section{Extrapolation formulas}
%===============================
To study the effects of the box size $L$ on the binding 
energy $B_N$ of an $N$-body system it is convenient to focus on 
the difference $\Delta B_N = B_N(L)-B_N^{\mathrm{free}}$ between
the finite volume and the free space binding energies.
In the limit $L\to\infty$ the finite volume binding energy $B_N(L)$ coincides with $B_N^{\mathrm{free}}$, and $\Delta B_N $ vanishes.

For an $s$-wave two-body bound state, the leading correction to $\Delta B_2$ 
in the limit of large box size was derived by L\"{u}scher \cite{Lucher, Lucher2}.
L\"{u}scher's formula is given by 
\be \label{Luscher}
  \Delta B_2 = \frac{6 \kappa_2 |\AS_2|^2}{ \mu_2 L} e^{-\kappa_2 L}\,,
\ee
where $\mu_2$ is the reduced mass, $\kappa_2 = \sqrt{ 2 \mu_2 B_2^{\mathrm{free}}}$ is the binding momentum, and $\AS_2$ is the dimensionless two-body asymptotic normalization coefficient (ANC). Here we use $\hbar=1$. 

%corrections
A box with periodic boundary conditions is equivalent to an infinite array of identical
boxes. Denoting an arbitrary cell as the "center" of the system ${\bf n}=(0,0,0)$,
we note the the leading correction \eqref{Luscher} is due
to summation over nearest neighbours,
i.e. it includes only the first layer of cubic volumes with $|{\bf n}|^2=1$. Summation over more distant layers -- second $|{\bf n}|^2=2$, third $|{\bf n}|^2=3$, etc. provides the sub-leading corrections \cite{KL:corrections}
\begin{align}
\label{Luscher:corr}
&\Delta B_2 =\\ 
&\frac{\kappa_2 |\AS_2|^2}{\mu_2 L} \left(\overbrace{6 e^{-\kappa_2 L}}^{|{\bf n}|^2=1}+\overbrace{\frac{12}{\sqrt{2}} e^{-\sqrt{2} \kappa_2 L}}^{|{\bf n}|^2=2}+\overbrace{\frac{8}{\sqrt{3}} e^{-\sqrt{3} \kappa_2 L}}^{|{\bf n}|^2=3}+\ldots\right)\,. \nonumber
\end{align}

%few-body systems
A similar approach can be used to analyze also $N$-body bound states. If the lowest $s$-wave threshold of the system is breakup into two subclusters, then treating these subclusters as point-like particles yields the leading correction \cite{KonLee18}
\be\label{Luscher:few-body}
  \Delta B_N = C_N \frac{6\kappa_N |\AS_N|^2}{\mu_N L} e^{-\kappa_N L}\,,
\ee
where $\mu_N$ is the reduced mass of the subclusters and $\kappa_N = \sqrt{2\mu_N (B^{\mathrm{free}}_N-B^{\mathrm{free}}_1-B^{\mathrm{free}}_2)}$ is the binding momentum calculated using the $N$-body binding energy $B^{\mathrm{free}}_N$ and the binding energies $B^{\mathrm{free}}_1$, $B^{\mathrm{free}}_2$ of subclusters $1$ and $2$. Here, $C_N$ stands for a combinatorial factor that counts the number of options to partition $N$ identical particles into two subclusters and $\AS_N$ denotes the ANC corresponding to the asymptotic part of the $N$-body wave function given by the inter-cluster coordinate. In principle, one should sum over corrections from all possible cluster configurations, however, the contribution of the lowest threshold is usually the dominant one. 

Considering here only the ground states of the $s$-wave nuclei $\rm ^2H$, $\rm ^3H$, and $\rm ^4He$, we note that the lowest separation thresholds are those associated with breakup into $(N-1)$-particle subcluster and one nucleon.    

%------------------------------------------------
\subsection{Asymptotic Normalization coefficient}
%------------------------------------------------
%defining ANCs
A crucial parameter in Eqs.~(\ref{Luscher}), (\ref{Luscher:corr}), and (\ref{Luscher:few-body}) is the asymptotic normalization coefficient (ANC) which describes the magnitude of the long-range part of the $N$-body bound state wave-function in the 
channel of two subclusters.

The ANC value for one nucleon separation can be extracted from the overlap function \cite{NolWir11}
\begin{align}\label{ANCoverlap}
  &f_{N~lj}^{J_{N-1} J_N}(r) =\\ \nn
  &\left \langle \left[\Psi_{N-1}^{J_{N-1}} \otimes \left[ 
      \chi_{1/2} \otimes Y_l(\hat{\pmb{r}})
      \right]_{j} \right]_{J_N}\right| \frac{\delta(r-{{\eta}}_{N-1})}{r^2}
  \left| \Psi_N^{J_N} \right \rangle\,,
\end{align}
where $\Psi_N^{J_N}$ and $\Psi_{N-1}^{J_{N-1}}$ are the $N$-body and the residual $(N-1)$-body wave functions with the total angular momenta $J_N$ and $J_{N-1}$, respectively.  $\pmb{\eta}_{N-1}=\left(\rv_N-\sum_1^{N-1}\rv_i/(N-1)\right)$ is the Jacobi coordinate corresponding to a distance between the last nucleon and the center of mass of the subcluster. In line with Eqs.~(\ref{Luscher}), (\ref{Luscher:corr}), and (\ref{Luscher:few-body}) we consider only $s$-wave breakup, consequently, $l=0$ and the total momentum $j$ of the last nucleon is given only by its spin component $\chi_{1/2}$. Isospin part is not displayed for readability.

Dropping the quantum numbers in $f_{N~lj}^{J_{N-1} J_N}(r)$, which in this work are uniquely related to one nucleon $s$-wave breakup thresholds of $\rm ^2 H$, $\rm ^3 H$, and $\rm ^4 He$, the asymptotic behavior of $f_N(r)$ is given by
\be \label{ANCasymp}
   f_{N}(r \rightarrow \infty) =  \AS_N \sqrt{2 \kappa_N} \frac{e^{-\kappa_N r}}{r}\,,
\ee 
where we assume no Coulomb interaction.

The ANC value can then be extracted from the asymptotic value at $r\to\infty$ of the function 
\be \label{ANCextract}
   \AS_N (r) = \frac{1}{\sqrt{2 \kappa_N}}r~e^{\kappa_N r} f_N(r)\,.
\ee
Note, that through this work we adopt the convention of dimensionless ANC \cite{HamPanRus17}. In experiment the ANC has usually units of ${\rm fm}^{-1/2}$ and it is related to our ANC by a factor of $\sqrt{2 \kappa_N}$.

%EXPERIMENTAL ANCs
For the two-body case, $\AS_2$ approaches unity in the limit of zero range interaction. The first finite range correction takes into account the effective range $\reff$ 
\be \label{ANCreff}
  |\AS_2|^2=\frac{1}{1-\kappa_2 \reff}\,.
\ee
Using experimental deuteron binding momentum and $NN$ spin-triplet effective range yields for the deuteron $\AS_2 = 1.2975$, which is in agreement with the experimental value $\AS_2 = 1.2902(65)$ \cite{Borbely85}. For triton the corresponding $s$-wave ANC value was extracted from an analysis of high-precision $d(d,p)t$ reaction data yielding $\AS_3\left({\rm ^3H}|n d\right)=2.19(2)$ \cite{Timofeyuk10}. Following Ref.~\cite{Timofeyuk10} we note that so far the $s$-wave $\AS_4\left({\rm ^4He}|pt\right)$ and $\AS_4\left({\rm ^4He}|n ^3{\rm He}\right)$ ANCs have not been firmly constrained. Experimentally suggested range is $5.34 - 8.34\fm^{-1/2}$ \cite{BlBoDo77} which roughly translates to dimensionless ANC values between $4.11$ and $6.42$.

%======================================
\section{Stochastic Variational Method}
%======================================

We solve the few-body Schr\"{o}dinger equation in free space using the stochastic variational method (SVM) \cite{VarSuz95,SuzVar98}. 
Here, the $N$-body wave function $\Psi$ is expanded in a correlated Gaussian basis
\begin{equation} \label{corrgauss}
   \Psi = \sum_i c_i~\hat{\mathcal{A}}
         \left\{{\rm exp}\left[-\frac{1}{2} \pmb{\eta}^T \tilde A_i \pmb{\eta}\right] 
                \chi^i_{S M_S} \xi^i_{I M_I} \right\}\,,
\end{equation}
where $\hat{\mathcal{A}}$ stands for the antisymmetrization operator over nucleons and $\pmb{\eta}^T=(\pmb{\eta}_1, ..., \pmb{\eta}_{N-1})$ is a set of Jacobi vectors. $\chi^i_{S M_S}$ and $\xi^i_{I M_I}$ denote the spin  and isospin part of the wave function, respectively, and $\tilde A_i$ is a positive-definite symmetric matrix that contains $N(N-1)/2$ stochastically selected parameters. 
The variational coefficients $c_i$ and the corresponding bound state energies are calculated by diagonalizing the Hamiltonian matrix, i.e. solving the generalized eigenvalue problem.
    
%---------------------------------------
\subsection{Periodic Boundary Condition}
%---------------------------------------
Putting the system in a box with periodic boundary conditions, one has to solve
the Schr\"odinger equation  
\be
  H_L \Psi_L = E_L \Psi_L\,,
\ee
where the subscript $L$ denotes the lattice. On the lattice, the wave function $\Psi_L$ is to 
obey the periodic boundary conditions,
\be\label{PBC}
\Psi_L(\rvec_1, \rvec_2, \dots) = \Psi_L(\rvec_1+\mvec_1 L, \rvec_2+ \mvec_2 L, \dots)\,,
\ee
for arbitrary integer trios $\{\mvec_1,\mvec_2,\ldots\ ; \mvec_i^T=(m_{ix}, m_{iy}, m_{iz}) \}$.
The periodic potential $V_L$ is given by
\be\label{poten}
    V_L(\rvec_1,\rvec_2,\ldots ) = \sum_{\mvec_1,\mvec_2,\ldots }
         V(\rvec_1+\mvec_1L,\rvec_2+\mvec_2L,\ldots )\,.
\ee  
For example, the $x$-axis component, $x_{ij}=x_i-x_j$, of the two-body potential becomes
$$
\exp[-\Lambda^2 x_{ij}^2/4] \longrightarrow \sum_q \exp[-\Lambda^2 (x_{ij}-qL)^2/4]\,,
$$
where in principle the sum over $q$ runs over all integers. In practice, due to the short-range nature of the interaction, far boxes are negligible and the sum is limited to $-N_{\mathrm{max}}\leq q \leq N_{\mathrm{max}}$.

Earlier works \cite{YinBlu13,EliBazBar20,DetSha21} used a single-particle framework which facilitates implementation of the periodic boundary condition defined in these coordinates. The wave function is written as a product of functions in the $x,y,z$ directions,
\be 
  G_L(\rvec)= G_{Lx}(\xvec)G_{Ly}(\yvec)G_{Lz}(\zvec)\,.
\ee
The $x$-component basis functions $G_{Lx}$ (same for $G_{Ly},G_{Lz}$) are
defined by a symmetric positive definite $N \times N$ matrices $A_x$, 
a positive definite diagonal matrix $B_x$, and a shift vector
$\dvec = (d_1, \dots d_N)$.
To satisfy the periodic boundary conditions, one sums over shifts of this function, 
\be \label{oldWF}
G_{Lx} = \sum_{\nvec_x} G_x(A_x,B_x,\dvec_x;\xvec-L\nvec_x)\,,
\ee
with
\be
 G_x = \exp\left[-\frac{1}{2} \xvec^T A_x \xvec - \frac{1}{2}(\xvec-\dvec_x)^T B_x (\xvec-\dvec_x)\right]
\,,
\ee
and ${\nvec_x} = (n_{1x}, n_{2x}, \ldots, n_{Nx})$, $n_{ix} \in \mathbb{Z}$\,.

%---------------------------------------
\subsubsection{Center of Mass Reduction}
%---------------------------------------

Working in single-particle coordinates has a few disadvantages compared to internal coordinates.  
First, the number of parameters defining a basis function, Eq.~(\ref{oldWF}), is larger, and thus more parameters have to be optimized. 
Second, to enforce the periodic boundary condition one has to sum over the nearest boxes, and this summation is repeated for each coordinate.
Finally, it is complicated to suppress a center of mass excitation, and therefore to converge to the ground state energy.

In order to address difficulties related to the center of mass we introduce different basis functions,
\be \label{newWF}
G_{Lx} = \sum_{\nvec_x} \tilde{G}_x(A_x;\xvec-L\nvec_x)\,,
\ee
with
\be \label{newWFstate}
\tilde G_x(A_x;\xvec) = \exp \left[-\frac{1}{2}\xvec^T A_x\xvec \right]\,,
\ee
where each function carries a set of $N(N-1)/2$ parameters $\{b_{i,j}\}_{i<j}$ which reflect the relative distance between each pair of particles, from which the $N\times N$ matrix $A_x$ is defined as \cite{VarSuz95}
\be \label{newWFcnd}
   \xvec^T A_x \xvec = \sum_{i<j} \frac{(x_i-x_j)^2}{b^2_{ij}}\,.
\ee

Since these basis functions are invariant under center of mass shifts, the summation in Eq.~(\ref{newWF}) can be done without the last parameter $n_{Nx}$, 
i.e. ${\nvec_x} = (n_{1x}, n_{2x}, \ldots, 0)$, $n_{ix} \in \mathbb{Z}$.
This observation leads to substantial reduction in calculation time. 
For the proof see appendix A. 

Although the basis functions \eqref{newWF} are a special subset of
the basis states \eqref{oldWF}  
used in former box SVM calculations \cite{YinBlu13,EliBazBar20,DetSha21}, 
the matrix elements cannot be naively calculated using the formulas of Ref.~\cite{YinBlu13} due to appearing of unwanted infinities.
The technique to handle these infinities is given in appendix B.

\begin{figure*}
 \centering
  \begin{tabular}{cc}
 \includegraphics[width=8.6cm]{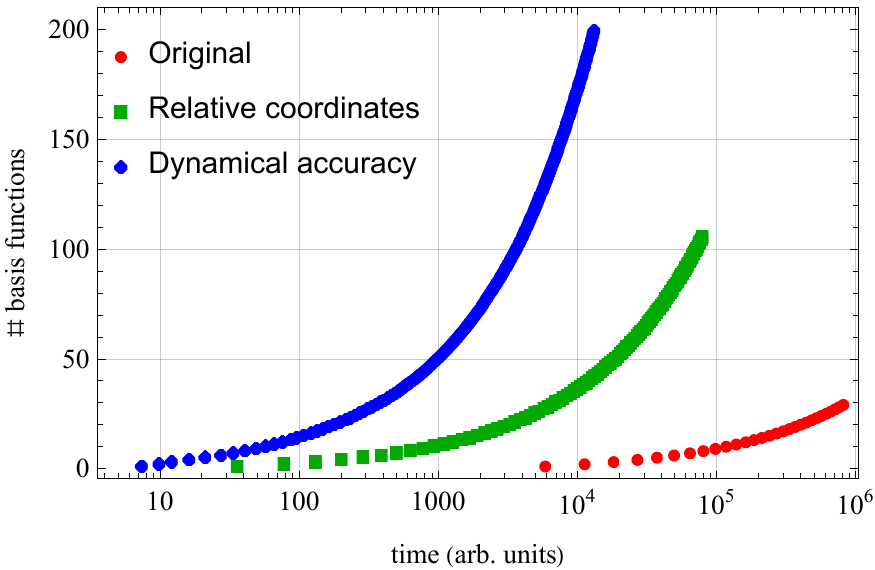}&\includegraphics[width=8.6cm]{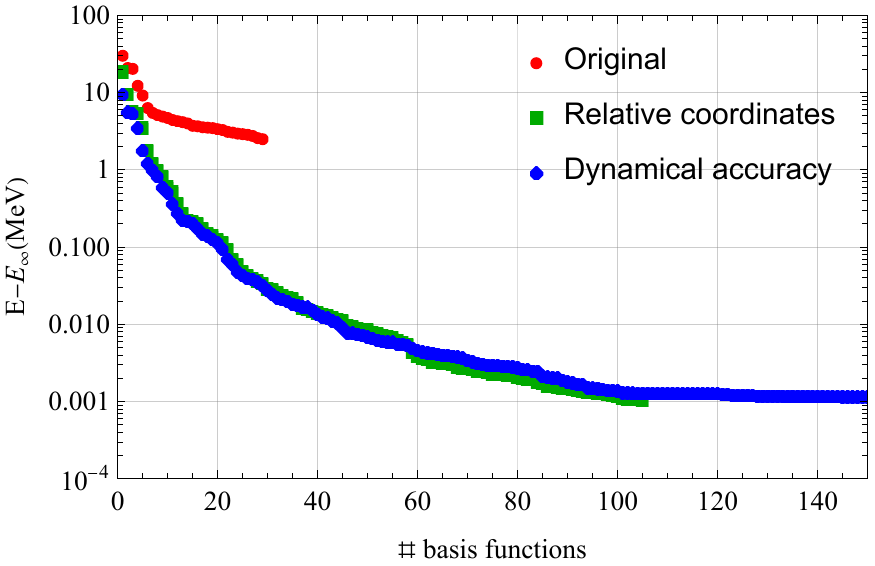}\\
 \end{tabular}
 \caption{Left panel: The number of basis functions added to the optimized basis as a function of the running time. Calculations were done for $\rm ^3H$, see text for details. 
  Shown are results for the original algorithm, with single-particle coordinates (red), for the modified algorithm using relative coordinates (green), and including also dynamical accuracy control (blue). Right panel: The difference between the binding energy estimation for a given basis size to the converged value. Calculations were done for $\rm ^3H$, see text for details.
 \label{fig:BSVMcode}}
\end{figure*}

%---------------------------------
\subsubsection{Dynamical Accuracy}
%---------------------------------
The sum over all possible shifts in particle coordinates, as well as in the potential, makes a calculation of the corresponding matrix elements computationally demanding -- it is the main bottleneck of the Box-SVM. However, both the wave function and the inter-particle interaction decay exponentially, and therefore the summation can be limited. In the original version of the Box-SVM code a single parameter, $N_{\mathrm{max}}$, controlled the maximal absolute value of $q$ appearing in Eq.~\eqref{poten} or any element of $\nvec_x$ entering Eq.~\eqref{oldWF}. 

The SVM optimizes parameters of the basis function to be added to the basis set in a stochastic way. Thus, each parameter is chosen from a large sample; this process involves hundreds of calculations of Hamiltonian and overlap matrix elements in a search for such a parametric set which significantly lowers the bound state energy. 
Consequently, most of the run time is spent on calculations in the selection process, while considerably less amount of time is invested to calculations with an already selected basis set. This allows further improvement in the Box-SVM algorithm, where calculations involved in the selection process can be done with lower accuracy, while highly accurate -- and computationally demanding -- calculations are done only with the selected basis.
\\[10pt] We optimize the Box-SVM algorithm in several ways:\\ 

First, one can use two different parameters to dictate the summation limits. 
The frequent and less accurate selection process is done with a lower summation limit, $N_{\mathrm{low}}$, while the rare and more accurate calculations with the chosen basis are done with a higher summation limit $N_{\mathrm{high}}$.

Second, a parameter for the desired precision could be defined, and then the summation is finished when the change in a calculated matrix element is less than the desired precision. 
Here also the required precision in the selection process can be less than the one of the final calculation. This optimization is referred to as "dynamical accuracy". 

Finally, a large basis that was trained for a system without any box, and thus can be achieved efficiency, can be used to perform the calculations in a box. 
This is most effective for gaining high accuracy for large boxes and for the four particles system which is hard to calculate otherwise.

%-----------------------
\subsubsection{Run Time}
%-----------------------

To check the effect of the aforementioned modifications we devised three versions of the Box-SVM code. 
The first version is the original one, using single-particle coordinates and summing shifts up to a constant $N_{\mathrm{max}}$ \cite{EliBazBar20}. 
The second version uses relative coordinates, thus separating center of mass motion. The last version has, in addition, dynamic accuracy control on the summation. 

As an example we calculate the triton binding energy using these versions of the code, and measure the number of optimized basis functions added to the basis and the estimations for the ground state energy as a function of the running time. 
The calculations were done for a box size of $L=5\fm$, using LO \nopieft with $\Lambda = 4\invfm$, and with the summation control parameters 
$N_{\mathrm{max}}=N_{\mathrm{high}} = 7$ and $N_{\mathrm{low}} = 4$.

The left panel of Fig.~\ref{fig:BSVMcode} shows the number of optimized basis functions found as a function of the running time for the three versions. The improvement of about an order of magnitude in run time for each code modification is clearly seen. 

The right panel of Fig.~\ref{fig:BSVMcode} compares the difference between the binding energy estimation for a given basis size to the converged value. The original code converges slowly, probably due to contamination from center of mass excitations. Both improved versions show faster convergence at a similar rate. 

%================
\section{Results}
%================

We calculate the binding energies of the light nuclei -- deuteron ($^2$H), triton ($^3$H) and $^4$He, in LO \nopi EFT. 
Calculations are done using the SVM in free space and in boxes of various sizes $L\in[2,70]\fm$, with periodic boundary conditions.

The deuteron and triton binding energies in free space were used to fit the EFT LECs and therefore are cutoff independent. The $^4$He binding energy is a prediction of the theory and thus has residual cutoff dependence. 
This results in a variation of about $6$ MeV in $^4$He binding energy between $\Lambda=1\invfm$ and $\Lambda=10\invfm$.

As mentioned above, two parameters govern the box shift to the binding energy -- the ANC and the relative binding momentum between the subclusters of the system. The most important subclusters are those with the lowest threshold. In our case, it is a separation of one nucleon. The relevant momentum values are $\kappa_2 = 0.232~{\rm fm^{-1}}$, $\kappa_3 = 0.449~{\rm fm^{-1}}$, and $\kappa_4 = 0.931 - 0.814~{\rm fm^{-1}}$ for cutoff values between $\Lambda=1\invfm$ and $\Lambda=10\invfm$.

\begin{figure*}[t]
 \centering
 \begin{tabular}{cc}
 \includegraphics[width=8.6 cm]{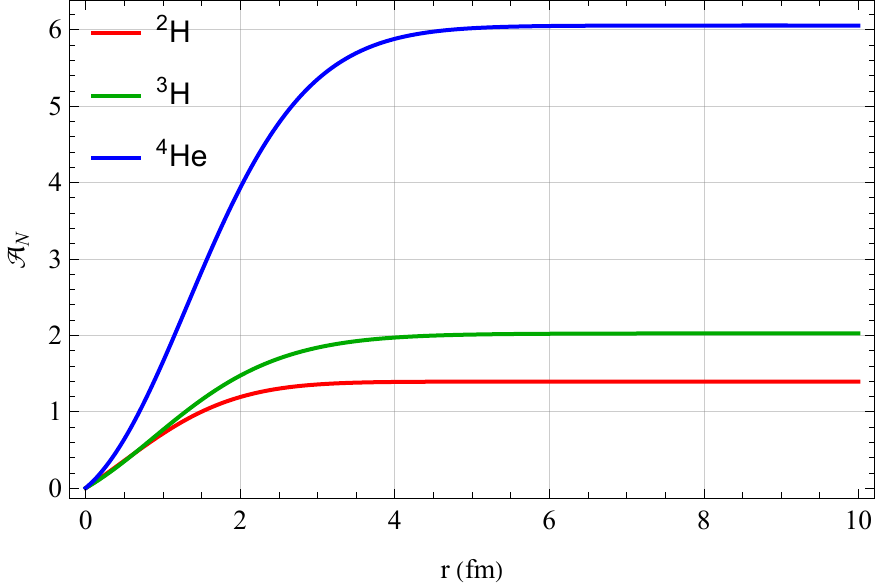}&\includegraphics[width=8.6 cm]{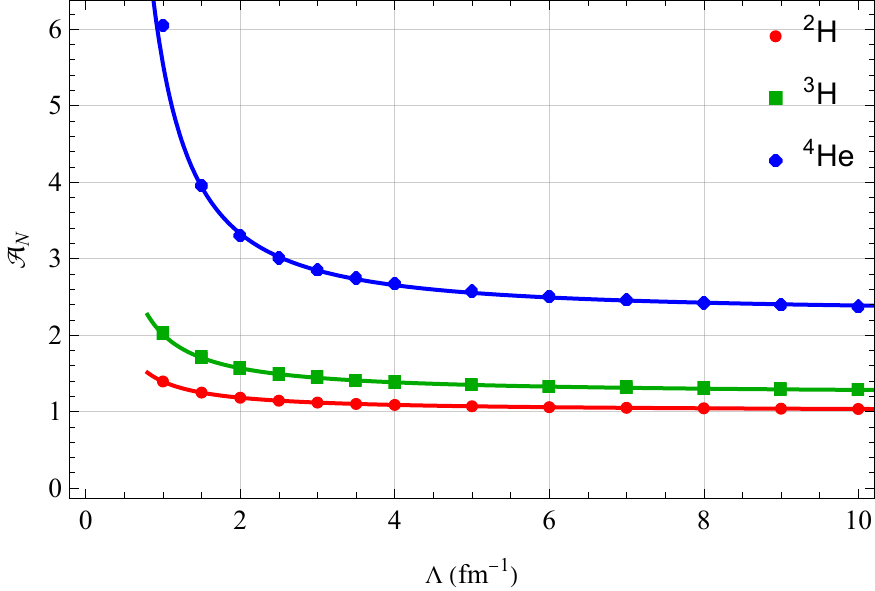}\\
 \end{tabular}
 \caption{Left panel: The nuclear ANCs $\AS_N$, calculated using Eq.~(\ref{ANCextract}), as a function of the distance $r$ between the last nucleon and the remaining subcluster. Shown are results for the one nucleon $s$-wave disassembly threshold of the deuteron (red), triton (green), and $^4$He (blue) ground states. 
 Right panel: The ANCs values extracted using the plateau in the overlap function, as a function of the cutoff. Shown also is the best polynomial fit (solid line), which is used to extract the value in the $\Lambda \longrightarrow \infty$ limit. 
 \label{fig:ancs}}
\end{figure*}

\subsection{Free space}
The ANCs were obtained from free space calculations using the overlap function $f(r)$ defined in Eq.~(\ref{ANCoverlap}). In Fig.~\ref{fig:ancs} (left panel) we show calculated $\AS_N$s as a function of the relative distance $r$ between the last nucleon and the remaining subclusters. The results are shown for deuteron, triton, and $^4$He, with $\Lambda=1\invfm$. Clearly, the calculated ANCs converge fast with increasing $r$ allowing us to accurately extract the corresponding asymptotic values. 

The right panel shows extracted $\AS_N$ values as a function of the momentum cutoff $\Lambda$. In order to obtain the corresponding zero-range $\Lambda\longrightarrow\infty$ limit we fit our results using the extrapolation formula 
\be
  \AS_N(\Lambda)=\AS_N(\infty) + \frac{\alpha}{\Lambda} + \frac{\beta}{\Lambda^2}\,.
\ee
The $\AS_N(\infty)$ as well as $\Lambda\approx1.25\invfm$ results, where experimental $NN$ effective ranges are roughly reproduced, are presented in Tab.~\ref{tbl:ANC}. For completeness, we compare the calculated ANCs to the corresponding experimental values.
\begin{table}
\begin{center}
\caption{The ANC for one nucleon separation in light nuclei.}
\label{tbl:ANC}
\vspace{0.3cm}
{\renewcommand{\arraystretch}{1.25}%
\begin{tabular}
{c@{\hspace{5mm}} c@{\hspace{5mm}} c@{\hspace{5mm}} c@{\hspace{5mm}}}
\hline \hline 
ANC & $\Lambda=1.25\invfm$ & $\Lambda\rightarrow\infty$ & Exp.\\
\hline 
$\AS_2$ & 1.30 & 1.00 & 1.2902(65)  \\
$\AS_3$ & 1.82 & 1.22 & 2.19(2)  \\
$\AS_4$ & 4.52 & 2.27 & 4.11-6.42  \\
\hline \hline 
\end{tabular}}
\end{center}
\end{table}

%Deuteron
For the deuteron, the ANC should converge to 1 in the zero range limit, and indeed this is the case. Moreover, for all but the smallest cutoff, the change in the ANC is described well by the effective range correction, Eq.~(\ref{ANCreff}).

%Triton
Similarly, for triton $\AS_3(\infty)$ we obtain a result close to 1. This is in agreement with Ref.~\cite{HamPanRus17} where the authors demonstrated that for unitary two-body interaction and in the zero-range limit the three-body ANC possesses a value very close to unity. We checked that as the spin-triplet interaction is tuned closer to unitarity, i.e. shallower deuteron pole, the extracted $\AS_3(\infty)$ converges toward 1.

%4He
Throughout this work, we disregard Coulomb interaction. Consequently, our four-body ANCs, extracted from $\bra p~t | \alpha \ket$ overlap function, are equivalent to 
the $\bra n ^3\rm{He}| \alpha \ket$ ANCs and should be compared to this experimental quantity.  

%Benchmarks, precision of free space ANCs
Calculation of ANCs was successfully tested with respect to the previous Faddeev-type work for phenomenological $NN$ interaction \cite{PayFri80}. Results will be reported elsewhere. We note that the extraction of two and three-body ANCs, done in this work, is highly accurate and their error is negligible. For four-body ANCs the error is of the order of $10^{-3}$. 

%-------------------------------
\subsection{Periodic Finite Box}
%-------------------------------
Equipped with the free space values for the binding energies, binding momenta $\kappa_N$, and the asymptotic normalization coefficients $\AS_N$, we move to the finite volume results. Using our upgraded Box-SVM code we calculate binding energies of $\rm ^2H$, $\rm ^3H$, and $\rm ^4 He$ for a wide range of box sizes.

\begin{figure*}
 \centering
   \begin{tabular}{cc}
 \includegraphics[width=8.6 cm]{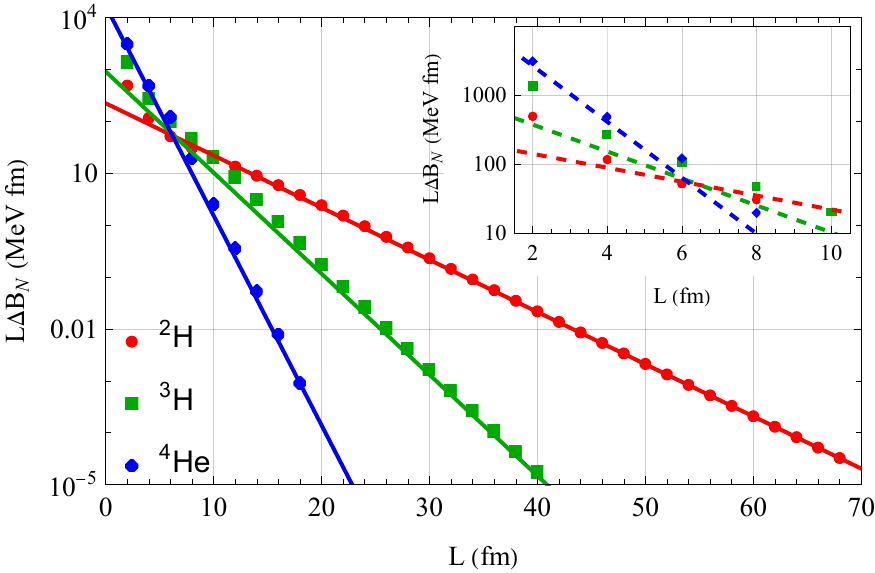}& \includegraphics[width=8.6 cm]{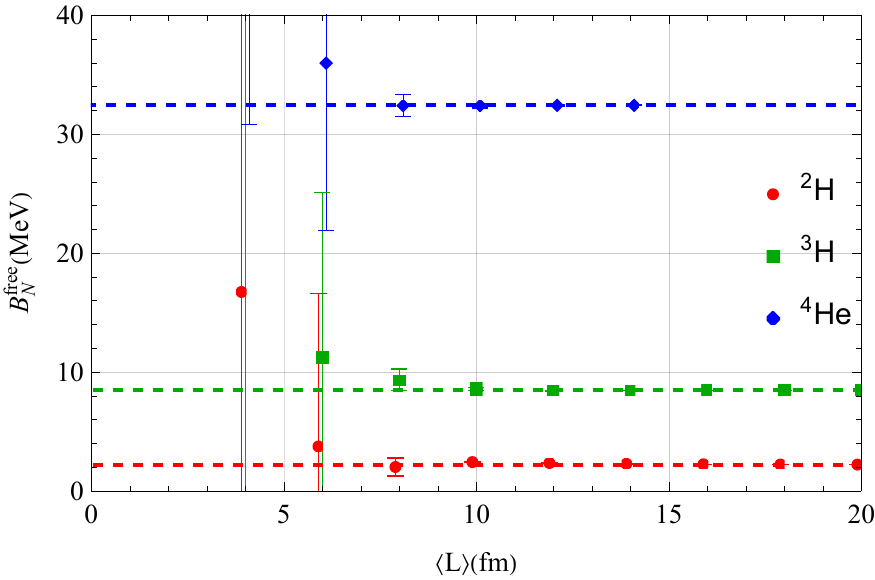}\\
 \end{tabular}
 \caption{Left Panel: The energy shift $\Delta B_N$ due to the box, multiplied by the box size $L$, as a function of $L$. Shown are results for deuteron (red), triton (green), and $^4$He (blue), using LO \nopieft with $\Lambda=1\invfm$. Points represent Box-SVM results. Lines are the predictions of L\"{u}scher, Eq.~(\ref{Luscher}) and the few-body formula, Eq.~(\ref{Luscher:few-body}) using free-space $\AS_N$ and $\kappa_N$ values. The subplot zooms in small boxes. Left Panel: The free-space binding energy values $B_{N}^{\rm free}$ extracted from the energies in a box, using these equations. Shown are results calculated from three adjacent box sizes, as a function of the middle box size. The free space binding energies are shown as a dotted line.
 \label{fig:dBCut1}}
\end{figure*}

The results for LO \nopieft with $\Lambda=1\invfm$ are shown in the left panel of  Fig.~\ref{fig:dBCut1}. Here, we depict energy shifts $\Delta B_N$, multiplied by the box size $L$, as a function of $L$. Presenting our results in a logarithmic scale reveals dominant $e^{-\kappa_N L}$ behavior which translates into linearly arranged finite volume results -- red (deuteron), green (triton), and blue ($\rm ^4He$) symbols. 

In the same plot, we show prediction of L\"uscher formula, Eq.~(\ref{Luscher}), using the relative momenta and ANC extracted from free space deuteron (red line), as well as its extension, Eq.~(\ref{Luscher:few-body}), for the triton (green line) and $^4$He (blue line). Clearly most of the finite volume effect is captured by these formulas. However, one should keep in mind that these are asymptotic formulas and thus have large deviations for small boxes. To emphasize this the subplot zooms in the small boxes $L \le 10$ fm, where these deviations are clearly seen.

In return, we can use Eqs.~(\ref{Luscher}) and (\ref{Luscher:few-body}) to fit calculated $B_N(L)$ energies, thus obtaining free-space $B_N^{\rm free}$ and $\AS_N$. To simulate the typical LQCD situation, we restrict ourselves to use only three adjacent box sizes. In the right panel of Fig.~\ref{fig:dBCut1} we present the resulting binding energies, as a function of the middle box size. The results from the smallest boxes, $L=\{2,4,6\}$ fm, deviate considerably from the free space values. To obtain free-space predictions with an accuracy better than 1 MeV, larger boxes, with $L\geq 6$ fm should be used. The results of this approach for $L=\{6, 8, 10\}$ fm are also presented in Tab.~\ref{tbl:Lusher} along side the corresponding free space results. Apparently, the extracted binding energies match reasonably well those calculated directly in free-space.

\begin{table}
\begin{center}
\caption{Free-space binding energies and one nucleon separation ANCs in light nuclei.
A comparison between free-space calculations and the L\"uscher formula results, using boxes with $L=\{6,8,10\}$ fm.}
\label{tbl:Lusher}
\vspace{0.3cm}
{\renewcommand{\arraystretch}{1.25}%
\begin{tabular}
{c@{\hspace{5mm}} c@{\hspace{5mm}} c@{\hspace{5mm}} c@{\hspace{5mm}} c@{\hspace{5mm}}}
\hline \hline 
      & \multicolumn{2}{c}{Free-space}         & \multicolumn{2}{c}{L\"uscher} \\
  $N$ & $B_N^{\rm free}$ [MeV] & ANC & $B_N^{\rm free}$ [MeV] & ANC\\
\hline 
  2 & 2.2246 & 1.40  & 2.0(8)  & 1.3(2) \\
  3 & 8.482  & 2.024 & 9.4(9)  & 2.2(1) \\
  4 & 32.48  & 6.00  & 32.4(9) & 4.9(1) \\
\hline \hline
\end{tabular}}
\end{center}
\end{table}

\begin{figure*}
 \centering
  \begin{tabular}{cc}
 \includegraphics[width=8.6 cm]{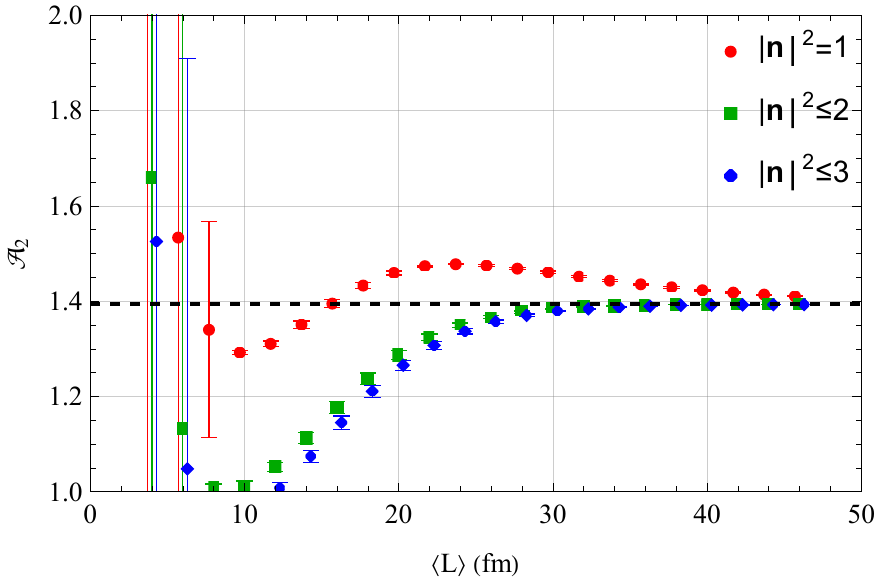}& \includegraphics[width=8.6 cm]{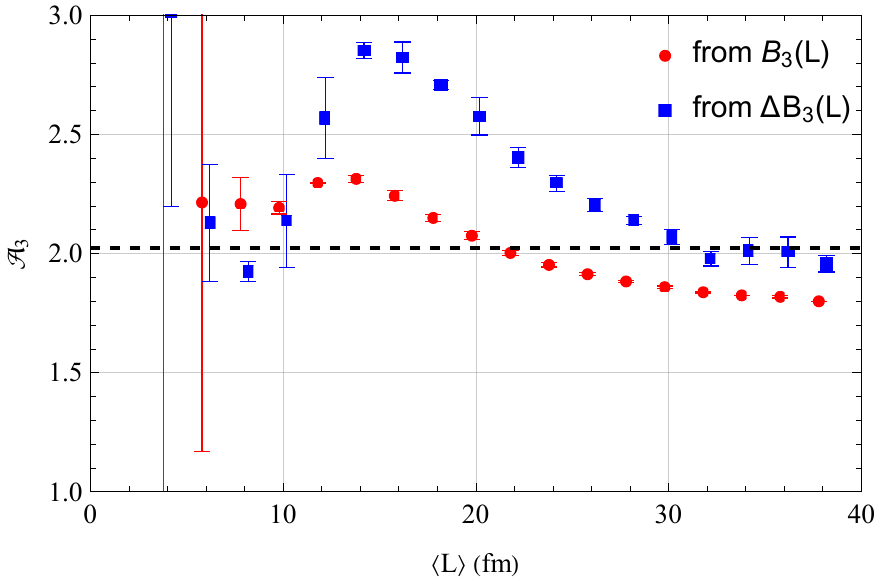}\\
 \end{tabular}
 \caption{The deuteron (left panel) and triton (right panel) ANC extracted from the energies in a box for $\Lambda=1\invfm$. Shown are results calculated from three adjacent box sizes, as a function of the middle box size. The free space ANCs are shown as a dotted line. Left Panel: The deuteron ANC extracted using L\"uscher formula with only nearest neighbours (red), as well as including also next terms (green and blue). Right Panel: The triton ANC extracted using Eq.~(\ref{Luscher:few-body}),  
 from fitting $B_3^{\rm free}$ and $\AS_3$ from calculated $B_3(L)$ (red), and those from fitting $\kappa_3$ and $\AS_3$ from calculated $\Delta B_3(L)$ (blue).}
 \label{fig:LANC23}
\end{figure*}

Larger boxes are needed for exact extraction of the deuteron ANC. We present in the left panel of Fig.~\ref{fig:LANC23} the deuteron ANC for $\Lambda=1\invfm$, as extracted from calculated $B_2$s for three adjacent box sizes using Eq.~(\ref{Luscher}). Summing over the nearest neighbors only, i.e. using the $|{\bf n}|^2=1$ term, one has to consider extremely large boxes to extract $\AS_2$ precisely. For example, accuracy of $\approx 2\%$ is obtained for size $L\gtrsim 40\fm.$ Taking another order in this series, i.e. summing also next to nearest neighbors $|{\bf n}|^2\leq 2$, improves the convergence, and now $L\gtrsim 26\fm$ gives the $2\%$ precision. Including further order $|{\bf n}|^2 \leq 3$ has only minor effect. 

Extraction of the triton ANC is more involved. Here, minor changes in the fitting procedure can change the result significantly. To demonstrate that we present in the right panel of Fig.~\ref{fig:LANC23} resulting $\rm ^3H$ ANC using Eq.~(\ref{Luscher:few-body}) and $\Lambda=1\invfm$. We compare two ways of extracting the parameters. The first one considers calculated $B_3(L)$ values and fits $B_3^{\rm free}$ and $\AS_3$, while the second one obtains $\AS_3$ fitting directly calculated $\Delta B_3(L)$. Clearly, the extracted ANCs exhibit $L$ dependence in the whole box size range considered in this work. Moreover, even for $L\geq 15$~fm, where the extracted ANCs begin to converge, it seems that the converged values of these methods differ.

For $\rm ^4 He$ the calculations start to be a bit more computationally involved. Corresponding $\kappa_4$ binding momentum is approximately twice as large than $\kappa_3$ which makes energy shift $\Delta B_4$ to decrease rather fast $\sim (\Delta B_3)^2$. Inspecting the left panel of Fig.~\ref{fig:dBCut1} reveals that considering $L > 16$~fm demands $\Delta B_4$ accuracy $< 10^{-5}$~MeV which remains at the moment beyond the scope of our numerical precision. For $\Lambda=1\invfm$ we observe deviation up to 25\% of extracted $\AS_4$ using Eq.~(\ref{Luscher:few-body}) from its free space value and, similarly to $\rm ^3H$ case, beginning of the convergence to the free space $\AS_4$ at $L \geq 10$~fm.

%======================
\subsection{Discussion}
%======================

Based on our results we weigh three different options on how to extract free space quantities from finite volume binding energies. While we will refer to our $\Lambda=1\invfm$ calculations, with a range close to the range of the nuclear force, one should bear in mind that discussed finite-volume effects might slightly differ in case of more sophisticated $NN$ interaction or in a future \emph{ab initio} LQCD calculations. Nevertheless, it is reasonable to believe that the main conclusions remain settled and this discussion will be useful concerning future $\rm ^2H$, $\rm ^3H$, and $\rm ^4He$ LQCD studies at physical $m_\pi$.

The first option relies on a calculation of $B_N(L)$ for large enough $L$ and a simple estimation of the free space binding energy as $B_N^{\rm free} \approx B_N(L)$. As shown in the left panel of Fig.~\ref{fig:dBCut1} this might leads to error of order of $\approx 1$~MeV at $L=10$~fm. Considering larger volumes yields more accurate results. However, performing LQCD calculations for such large $L$ is at the moment highly difficult. Furthermore, one should calculate $B_N(L)$ for several values of $L$ to properly assess the error.

The second option represents an application of L\"{u}scher formula or its few-body extension. In this case, we identify as the main obstacle the relatively large finite volumes for which such an approach starts to be applicable. Closely inspecting the left panel of Fig.~\ref{fig:dBCut1}, calculated $\Delta B_N(L)$ deviate from the expected $~e^{-\kappa_N L}$ behaviour for $L\leq 10$~fm. While the discrepancy might seem minor on a logarithmic scale, it can be observed more severely in Fig.~\ref{fig:LANC23} in terms of extracted ANCs. Considering subleading contributions from neighbouring boxes $|{\bf n}|^2 \leq 3$ do not improve description for $L\leq 10$~fm. Possibly, one might introduce more involved correction taking into account the interplay between finite volume and interaction range, however, this is beyond the scope of this work. As demonstrated in the right panel of Fig.~\ref{fig:dBCut1} trying to extract free-space binding energies, fitting calculated $B_N(L)$s via L\"{u}scher formulas, leads to rather large uncertainties once $B_N(L)$ results below $L \approx 6$ fm are considered. 
For the two-nucleon case, our results agree qualitatively with those of former works. Ref.~\cite{Beane04} found that for accurate extraction of the deuteron binding energy lattices of sizes $L \gtrsim 10$ fm are needed. An accuracy of $1^{\circ}$ in the $^1S_0$ channel phase shifts requires $L \gtrsim 5$ \cite{KL:corrections3}. 
Better extrapolation is possible using a few boosted states \cite{Bour11,Davoudi11,Briceno13}.
In conclusion, an extraction of $B_N^{\rm free}$ via Eqs.~(\ref{Luscher}) and (\ref{Luscher:few-body}) in a region of small finite volumes should be used with caution.

The third option is to follow an EFT approach. For example in \nopi EFT, LECs are fitted to finite volume binding energies. Free space quantities are then predicted from the same theory solving the conventional few-body Schr\"{o}dinger equation \cite{EliBazBar20}. At LO \nopieft there are in total three LECs, two two-body and one three-body, which might be constrained by $B_N(L)$ of dineutron, deuteron, and triton. Importantly, one requires only one set of the corresponding finite volume binding energies, not necessarily for the same $L$. Chosen $L$ should take into account the numerical precision of a finite-volume few-body bound state method which is employed in a fit of LECs. 

On the other hand, any EFT prediction is always burdened with a truncation error which must be properly accessed. For example, LO free space $\kappa_4$ binding momentum varies with $\Lambda$ as much as $\sim 20 \%$ and, as shown in Fig.~\ref{fig:ancs}, even stronger $\Lambda$ dependence can be seen for $\AS_N$. The same source of error must be also expected for quantities that have been fixed in a finite volume and LO \nopieft was used as a tool to obtain corresponding free space predictions. While one can introduce rough estimates of this error, only proper inclusion of next-to-leading order (NLO) contributions provides accurate information. Within this regard, it is highly important to have access to more LQCD data, which might be either bound or excited states, as well as calculations with different $L$.

%===================
\section{Conclusion}
%===================

We explored the validity of the L\"{u}scher formula \cite{Lucher,Lucher2} and its few-body extension \cite{KonLee18} for $s$-shell nuclei $\rm ^2 H$, $\rm ^3H$, and $\rm ^4He$ at physical $m_\pi$ within LO \nopieft. Introducing a few upgrades into our Box-SVM 
algorithm we were able to significantly increase the precision of the few-body solution of the Schr\"{o}dinger equation in finite volume with periodic boundary conditions. Using the Box-SVM and its free-space equivalent we performed accurate calculations of the aforementioned systems in free space and finite volumes for a wide range of box sizes $L$. Equipped with calculated free-space asymptotic normalization coefficients (ANC), binding energies, and binding momenta, we used L\"{u}scher's approach to predict binding energy shifts $\Delta B_N(L)$ induced by the finite volume. This prediction was then critically analyzed with respect to $\Delta B_N(L)$ values obtained directly from the solution of finite volume Schr\"{o}dinger equation.

For \nopieft momentum cutoff $\Lambda =1 \invfm$, which is somewhat close to the range of nuclear force, these formulas describe reasonably well $\Delta B_N(L)$ energy shifts for $L\geq 10$~fm. With $L$ getting smaller we observe increasing discrepancy with respect to Box-SVM predictions. This is likely caused by an interplay between nuclear interaction range and finite-volume effects not accounted for by the formulas. Following the opposite direction, we tried to extract free space quantities from calculated Box-SVM energies $B_N(L)$ fitting the formulas. We conclude that binding energies can be obtained quite precisely, in particular, considering $B_N(L)$ values for $L \geq 6$~fm.
On the other hand, extraction of ANC for $\rm ^2H$ requires extremely large boxes $L>20$~fm and for $\rm ^3H$, $\rm ^4He$ we observe non-negligible $L$ dependence even for larger boxes. 

Nuclear LQCD calculations at the physical point will be likely performed at a box size smaller than $10$~fm, and extrapolation to the physical free space results will be needed. 
In the case $L \leq 6$ fm one should be aware of the inaccuracy of a L\"{u}scher like few-body formula caused by not properly separated scales -- nuclear and box sizes. Instead, one can use the \nopieft approach fitted directly to finite volume LQCD data \cite{EliBazBar20}. Here, systematic inclusion of higher orders of the theory might provide an accurate prediction of free-space quantities even from LQCD results at small $L$.

In the future, it would be interesting to extend our Box-SVM code to account for non-zero orbital momenta. This would allow studying $L$ dependence in few-body nuclear systems using realistic $NN$ interaction with non-central forces. Another topical task is to explore the insertion of NLO \nopieft terms fitting the corresponding LECs to LQCD finite volume data. While currently there are no available results at the physical point, LQCD data at heavier $m_\pi$ might serve as the starting point.

%========================
\section*{ACKNOWLEDGMENT}
%========================
The work of RY, MS, and NB was supported by the Pazy Foundation and by the Israel Science Foundation grant 1086/21.

%========
\appendix
%========
%------------------------------------------------
\section{The periodicity of a new basis function}
%------------------------------------------------
Here, we show that the basis function $G_{Lx}$, introduced in Eqs.~(\ref{newWF}-\ref{newWFcnd}), obeys the periodic boundary conditions, Eq.~(\ref{PBC}), even without summing over shifts of the last particle.

Since the basis function is defined only by the relative distances $x_i-x_j$ between particles $i$ and $j$, it is symmetric to a shift of all particles together,
\be\label{CC}
\tilde{G}_{x}(A_x;\xvec) = \tilde{G}_{x}(A_x;\xvec+{\bf C})\,,
\ee
for ${\bf C}^T = (C,C,\ldots,C)$.
The $G_{Lx}$ contains summation over all possible shifts of each particle except the last, therefore it obeys 
\be\label{sym}
G_{Lx}({\bf x}) = G_{Lx}({\bf x}+L{\bf n}) \quad \forall {\bf n}: n_N = 0\,.
\ee

For a general shift $L \nvec$ the required symmetry, imposed in Eq.~(\ref{PBC}), can be easily verified combining Eqs.~(\ref{CC})~and~(\ref{sym}).
Defining ${\bf b} = L{\bf n} - {\bf C}$ with ${\bf C}^T = (Ln_N,Ln_N,\ldots,Ln_N)$, fulfills  $b_N=0$, therefore
\begin{align}
G_{Lx}({\bf x}) & = G_{Lx}({\bf x} + {\bf C}) = \nn\\ 
                & = G_{Lx}({\bf x} + {\bf C} + {\bf b}) = \nn\\
                & = G_{Lx}({\bf x} + L{\bf n})\,,
\end{align}
which stands for the desired symmetry.

%------------------------------------
\section{Matrix elements calculation}
%------------------------------------
The SVM requires calculations of the Hamiltonian and the overlap matrix elements between basis functions. These matrix elements include a common factor, 
\be 
O=\sqrt{\frac{(2\pi)^N}{\det(A+A'+B+B')}}\,.
\ee 
For the basis functions \eqref{newWFstate} $B=B'=0$ and the $A+A'$ determinant is zero. This is because these basis functions do not define any constraint on the center of mass, while we do integrate over all coordinates. To cure this deficiency one can multiply the basis function by another Gaussian function, 
\be
G({\bf r}) \longrightarrow G({\bf r}) \exp\left[-\frac{1}{2}\lambda r^2_{CM}\right]\,,
\ee
where $\rvec_{CM}={\bf \eta}_N$ is the center of mass position, and $\lambda \longrightarrow 0^+$. Since at this limit we return to the original basis function, $\lim_{\lambda \rightarrow 0^+} O = O$. The determinant can be calculated now easily by transforming to the Jacobi coordinates to get 
\be
det\left[U_J(A+A')U_J^T\right]\times \int d\rvec_{CM} \exp(-\lambda r^2_{CM})\,,
\ee
where $U_J$ is a ($N \times N-1$) transformation matrix to Jacobi coordinates without the center of mass axis.  Therefore the determinant act only over the first $N-1$ axes in those coordinate system. The last integral appears now in the Hamiltonian and the overlap matrix elements and thus can be contracted safely.

%==========================

\end{document}